\documentclass[letter]{aa} 
\usepackage{graphicx}
\usepackage{txfonts}
\begin{document}
   \title{The       comet        \object{17P/Holmes 2007}             outburst:
   the      early            motion         of     the       outburst material.
   \thanks{Based on observations taken at the Wendelstein $0.8\,$m telescope.}}


   \author{M. Montalto
          \inst{1},
          A. Riffeser
          \inst{2},
          U. Hopp
          \inst{1,2},
          S. Wilke
          \inst{2},
          G. Carraro
          \inst{3}
          }

   \authorrunning{}

   \offprints{M. Montalto}

   \institute{MPE, Scheiner Str. 1, D 81679 Muenchen, Germany.\\
              \email{marco.montalto@unipd.it}
    \and
              Universitaetssternwarte  Muenchen,   Scheiner  Str. 1,  D 81679
              Muenchen, Germany.
    \and
             ESO, Santiago
             }

   \date{}


  \abstract
   {On October 24, 2007 the periodic comet \object{17P/Holmes}  underwent
    an astonishing outburst that increased its apparent  total     brightness
    from magnitude $V\sim17$ up to $V\sim2.5$  in  roughly two days.  In this
    contribution we report on Wendelstein $0.8\,$m telescope (WST) photometric
    observations  of  the   early   evolution    stages    of the  outburst.}
   {We   studied   the evolution of the                structure   morphology,
    its   kinematic,  and    provided  an estimate of the ejected dust mass.}
   {We  analized   $126$  images of the comet in the $BVRI$ photometric bands
    spread      between     26   October,   2007   and  November   20,   2007.
    The   bright    comet    core  appeared well separated from that one of a
    quickly  expanding dust cloud in all the data, and the bulk of the latter
    was contained in the field of view of our instrument during the days soon
    after   the   outburst, allowing precise estimates both of the separation
    velocities        of        the       two         luminous    baricenters,
    and        of           the            expansion    velocity  of the dust
    cloud. The ejected   dust  mass   was   derived    on     the   base   of
    differential photometry on background stars occulted by the moving cloud.}
   {The two   cores      were        moving         apart       from     each
    other        at        a            relative       projected     constant
    velocity of $(9.87\,\pm\,0.07)\,$arcsec/day ($0.135\,\pm\,0.001$ km/sec).
    In               the           inner          regions       of        the
    dust              cloud              we         observed     a     linear
    increase in size at  a mean constant velocity   of $(14.6\,\pm\,0.3)\,$
    arcsec/day                                ($0.200\,\pm\,0.004\,$ km/sec).
    Evidence        of          a   radial  velocity                gradient
    in      the         expanding       cloud       was      also       found.
    Our     estimate       for  the expanding coma's mass was of the order of
    $10^{-2}-1\,$ comet's mass    implying a significant disintegration event.
    }
   {
    We   interpreted our observations in the context of an explosive scenario
    which   was   more   probably   originated  by  some internal instability
    processes,        rather    than    an   impact with an asteroidal  body.
    Due to the peculiar characteristics of this  event, further  observations
    and investigations are necessary in order   to  enlight the nature of the
    physical           processes          that       determined            it.
   }

   \keywords{Comets: individual: \object{17P/Holmes} -- Solar System: general --
              }

   \maketitle
%

\section{Introduction}
\label{sect:introduction}
In this    Letter,    we     concentrate on the analysis of the outburst
of   comet     \object{17P/Holmes},     occurred   on   October 24, 2007
(Santana 2007). We  provide an overview analysis of the early  evolution
phases of the surrounding  cometary environment, in  the  period between
October   $26$,   $2007$     and   November   $20$, $2007$   considering
the global structure of the expanding material, its kinematic properties
and inferring the ejected dust mass.  Comet    brightenings  are    well
documented in the  literature, and in  general  they   can be associated
with  either evident  cometary  nuclear fragmentation or not (see e.g. Sekanina
et al. 2002   and  references therein).   Despite    their    recurrence,
the  magnitude   and  the characteristics of   the   outbursts occurring
to comet \object{17P/Holmes} largely   justify       its    longstanding
reputation in the annals of astronomy (Barnard 1896). The periodic comet
\object{17P/Holmes}  was  discovered  on November 6, 1892             by
E. Holmes   in  London,  during an   outstanding   brightness   increase,
followed       by         another similar event   on    January 16, 1893.
When        discovered    by      Mr. Holmes        the           comet
was         around         $5$      months      past         perihelion
                                                 $(\rm T\,=\,$June 13).
On       October       24,      2007            around $6$ months after
the last perihelion   passage                       $(\rm T\,=\,$May 4,
2007)                             the                             comet
underwent          a          similar              phenomenon        to
those observed more than   one               hundred              years
ago. The heliocentric distance of
the           object        at        the      time    of the two major
events was around            $\rm \Delta_{sun}\,=\,$2.39 AU         and
$\rm \Delta_{sun}\,=\,$2.44 AU   in  1892    and      2007 respectively,
and                 the                    orbital inclination remained
substantially            unaltered        during                  this
period                                      ($\rm i\,\sim\,20^{\circ}$).
This  Letter   is    structured    in    the    following    way:    in
Sect.~\ref{sect:observations},  we  present     the   observations   we
acquired, and the data reduction method. In Sect.~\ref{sect:morphology},
we discuss the    morphological     evolution     of     the  expanding
structure. In  Sect.~\ref{sect:cinematic}, we calculated the separation
velocities  between the comet    and    the   center of the dust cloud,
as well as the expansion velocity   of     the                    cloud.
In Sect.~\ref{sect:mass},   we  estimated   the   ejected   dust   mass.
Finally in Sect.~\ref{sect:conclusions},  we  sum   up     our  results.


\section{Observations and reduction of the data}
\label{sect:observations}
The        observations         were          acquired                  on
October      26/28/29/31, 2007  and November 2/5/20, 2007. A  total number
of   $126$    images were analized in  the    $BVRI$     filters.       An
overview          of           the         observations    is   shown   in
Tab.~\ref{Table:observations}.
The                                                                 images
were      acquired     using   the   MONICA CCD  camera, at the Cassegrain
focus of the 0.8$\,$m     Wendelstein telescope,  with    a          scale
of $0.5\,$arcsec/pix, and a field  of  view of around $8\,$x$\,8\,$ arcmin.
The  pre-reduction   process was done using  standard  reduction  software
(MUPIPE)    specifically   developed at  the  Munich  Observatory for  the
MONICA          CCD        camera              (G\"ossl \& Riffeser 2002).
The     weather conditions were in general clear for all   our  observing
nights          with          the           exception         of night 26,
which                was                 cloudy. Otherwise we were able to
acquire $1$ image in the $R$ band  during that night,        and to use it
for                                the                            analysis
of              the         kinematic properties of the cloud described in
Sect.~\ref{sect:cinematic}.    The   exposure times were comprised between
$40$              and           $200$       sec     for the whole dataset.

\begin{table}
\caption{Number  of  images of    \object{17P/Holmes}   taken   during
the   different         observing            nights        in      the
$BVRI$    filters.}
\label{Table:observations}
\centering
\begin{tabular}{c c c c c}
\hline\hline
DATE & $B$ & $V$ & $R$ & $I$ \\
\hline
 26/10/2007 & -  & - & $1$ & - \\
 28/10/2007 & $6$  & $6$ & $9$ & $11$\\
 29/10/2007 & $10$ & $5$  & $10$  & $10$ \\
 31/10/2007 & $10$ & - & $12$ & $13$ \\
 02/11/2007 & $4$ & $3$ & $5$  & - \\
 05/11/2007 & - & - & $5$ & - \\
 20/11/2007 & - & - & $6$ & - \\
\hline
\end{tabular}
\end{table}

During     night       28,     the photometric conditions were good and stable
over the whole night, and         we         thus acquired  some images of the
Landolt                standard                  field                  PG0918
in                    all                      our                    filters.
We obtained $2$  images in the $B$, $V$, $R$ bands,  and $4$ images in the $I$
band.
These                                    data                             were
reduced       exactly     in    the     same      manner  as   the  scientific
observations      (see       also       Sect.~\ref{sect:mass}). The magnitudes
of          the          standard         stars   were reported to $1$ airmass
and $1$ arcsec, and    finally    calibrated          to       the     Landolt
photometric     system,        thanks      to     the     Stetson  Photometric
Standard Fields~\footnote{http://www3.cadc-ccda.hia-iha.nrc-cnrc.gc.ca/community/STETSON/standards/}
data            of      PG0918.      We     found a total of $26$ common stars
between     our      catalog      and     that    one     provided by Stetson,
and     obtained the following calibration   equation   for   the $R$    band
against the $V-R$ color:

\begin{displaymath}
R\,-\,r\,=\,-0.07\,(\pm\,0.02)\,(V\,-\,R)\,+\,23.18\,(\pm\,0.01)
\end{displaymath}

The    $\rm RMS$        residual         of          the     fit         with
respect  to the best fitting least square linear    model was $\sim0.02\,$mag.
This        result        gives    an      idea of our photometric  precision
during that night, and will be useful   in    Sect.~\ref{sect:mass}, in    the
discussion of the ejected mass estimate.

%
%


\section{The evolution of the morphological structure}
\label{sect:morphology}

\begin{figure}
   \centering
   \includegraphics[width=0.8\columnwidth]{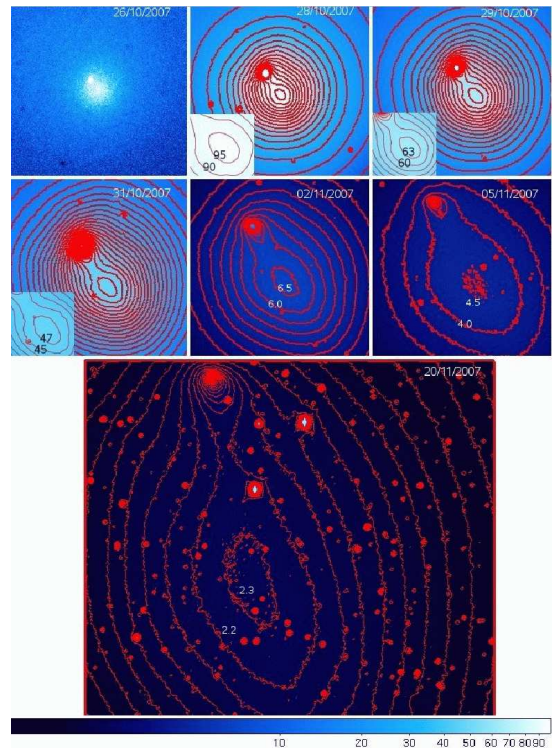}
   \caption{Snapshot of $R$ band  images acquired during each night of our
            observing     run,       showing        the   evolution     of
            comet        \object{17P/Holmes}'   morphological   structure.
            Each           image       has      been       scaled  to  the
            same exposure time, airmass,                intensity   range,
                                        and
            displayed on a logarithmic color scale, codified by the bottom
            color bar. The top $6$ images  are    zoomed  inside the inner
            regions    of     the      object, and have a field of view of
            $2.2\,$x$\,2.2$                  arcmin,               whereas
            the       last        one       of       $8\,$x$\,8$   arcmin.
            Two       cores are visible in all the    images,    the comet
            nucleus (on the upper left side)    and     that      one   of
            the coma,                        rapidly            separating
            during          the          observing                 period.
            The      values           reported            on           the
            coma's          innermost          isophotes    of        each
            panel (or in the close-up views of 28/29/31 nights)  put    in
            evidence   the   surface brightness decrement of the coma  due
            to its rapid expansion.
           }
              \label{Fig:morphology}%
    \end{figure}

In     order to show the  evolution of the morphological structure of
the object,     we       selected    for each night in our dataset one
representative image       in     the     $R$  band. These images were
scaled    to    the    same exposure time, airmass and intensity range,
in order to properly sample the   surface brightness of the coma on the whole
dataset. The lowest intensity level  was    chosen close to the lowest
counts        we        got         on          the  last image of our
dataset,       while         the         highest      close         to
the coma's peak intensity of the image taken   on  October  28,  2007.
We did not process in this way the image acquired on October 26, 2007,
because                  it             was        too          noisy.
The images were then normalized to the lowest intensity level. Finally, we
displayed        the        images       codified   on  a  logarithmic
color scale, as shown in Fig.~\ref{Fig:morphology}. We overplotted  to each panel
the isophote intensities levels, to highlight the structure of the inner
part of the cloud.        The           most               interesting
feature of Fig.~\ref{Fig:morphology} is the presence of two bright
cores            (the comet core beeing towards  the upper left side).
These    two    cores      were                           increasingly
separating              during the   observing    period and  appeared
elongated          towards       each    other, indicating an exchange
of mass between them.    Moreover the expansion of the global structure
is evident from the sequence of images. We also displayed the intensity value of  the closest
isophote to the cloud's center, and of        the adjacent one in the
outwards direction. The exact values given by the isophotes depends on the adopted number of
isophotes in each panel, which determines also the intensity step between
them. We set the number of isophotes in order to provide approximately
the same sampling of the inner regions of  the cloud. Thus, the quoted
numbers    must    be    considered    as   indicative of the relative
surface brightnesses of the cloud in the different nights. The intensity
of the innermost isophote of night $28/11/2007$ was around $2$ orders of
magnitude larger with respect to the correpondent isophote's intensity of night
$20/11/2007$, and the intensity step between the isophotes was also $1$
order of magnitude larger in the first case with respect to the second.
We                              underline                         that
there is no way to significantly change these conclusions adopting different
isophotal       mapping           criteria, as we accurately verified.
Actually, both these results are the consequence of the rapid expansion
of the dust cloud, which determines a steady decrease of its surface
brightness, and       a more homogeneous light distribution in the more
evolved, and less concentrated phases.
\\

%


\section{Kinematic of the expanding materials}
\label{sect:cinematic}

\begin{figure}
   \centering
   \includegraphics[width=0.24\textwidth]{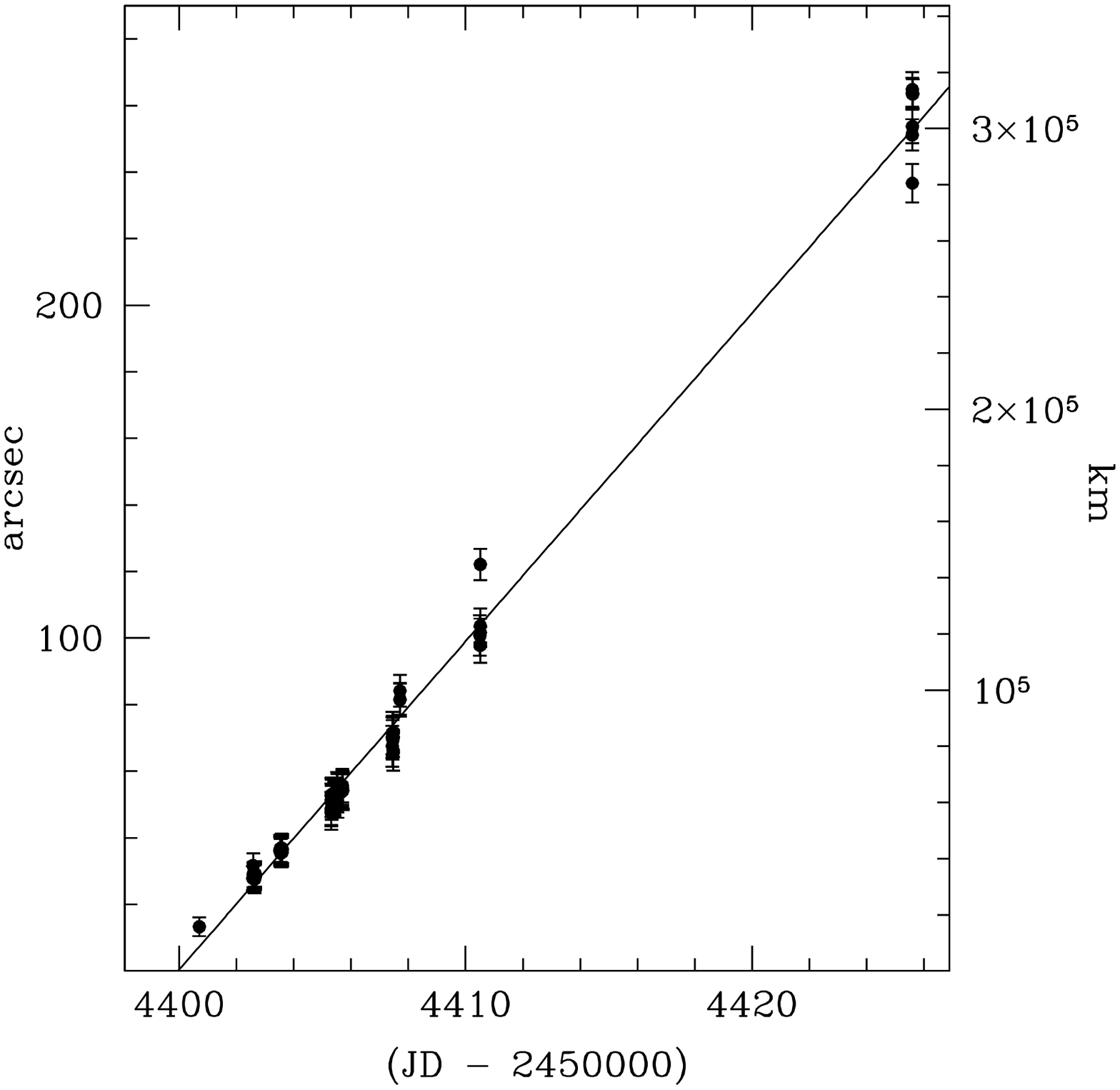}
   \includegraphics[width=0.24\textwidth]{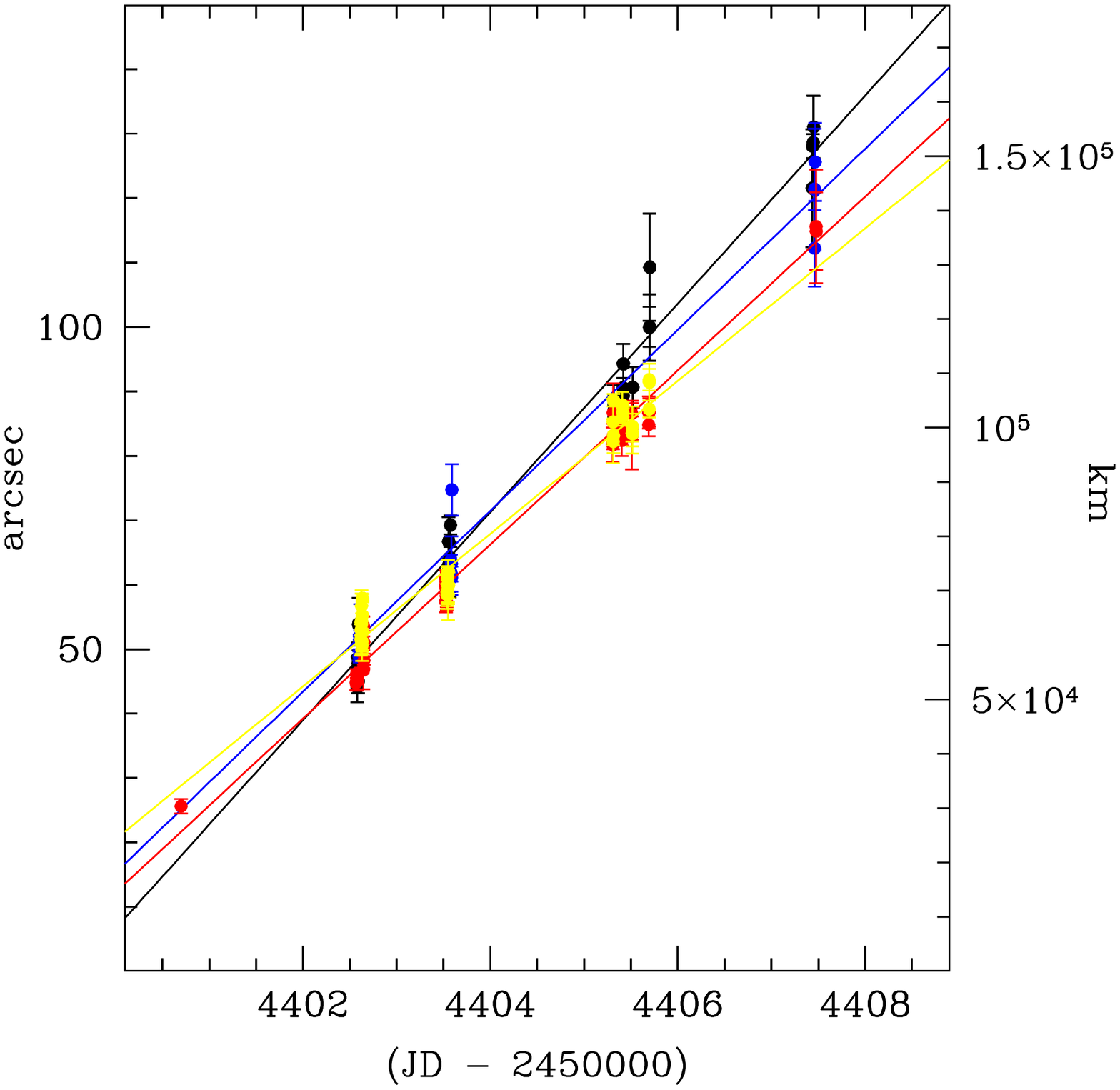}
   \caption{Left: angular     projected     distance         between      the
            centroid of the  two bright baricenters, for all the images in our
            dataset,      against           the         $\rm JD$          of
            the   observations. The continuous line denotes the best fit linear
            model. On the left hand y-axis values  are in   arcsec,  whereas
            on the right hand y-axis values are in km.
            Right:     $\rm \sigma$    of     the       expanding       cloud
            gaussian core      against $\rm JD$. Different  colors indicate
            measurements obtained in different bands: $B$ in  black,  $V$  in
            blue,  $R$  in  red  and  $I$ in yellow. The best fit  linear
            models in each band are represented by the continuous lines.
            }
              \label{Fig:velocities}%
    \end{figure}

In   this   Section  we derived the separation velocity of the two bright
baricenters, and     the     dust    cloud's      expansion velocity. The
comet's   center    appeared in all our images as a point-like source, thus
the peak's  position was  derived with a gaussian    fitting    algorithm,
as done for stellar objects. The light distribution of the dust cloud was
extended over a much larger area and was not gaussian overall, but we found
that the inner part could be always represented by a gaussian.
We   thus   fit     a        bidimensional
gaussian  using the light distributions along the x and y axis. An initial
guess for the centroid was provided with a maximum finding algorithm run
in a small region around the peak ($\sim\,30\,$arcsec). Then a least square
fit provided the refined position for the centroid,
and the $\sigma_x$  and $\sigma_y$ of        the
fitting         gaussians. In    Fig.~\ref{Fig:velocities}  (left panel)
we plotted the two bright cores' projected distances against the Julian Date
(JD) of the observations.           The       correction  for the Earth's
variable distance has been taken into  account,   although    negligible.
We obtained a uniform increasing  separation  with a mean  velocity    of
$(9.87\,\pm\,0.07)$ arcsec/day              ($0.135\,\pm\,0.001$ km/sec).
The uncertainties          are    dominated by the errors in the diffuse
cloud   centroids.  In   the  right   panel of Fig.~\ref{Fig:velocities},
we reported the  $\sigma$ of the diffuse clouds (actually the mean of the
$\sigma_x$          and         $\sigma_y$)    as a   function of the JD.
We      did   not   include   in this Figure      the                 last
two nights,                   because       the cloud was too expanded.
Two images acquired on night $2/11/2007$ in the $R$ band were also excluded because
of the large offset   of the  cloud's   centroid      with respect to the
image's        center. The errorbars in this Figure are calculated as the
difference     of     the     $\sigma$    along   the   x    and y  axis.
The  different    colors    in    the     figure    codify   observations
obtained  in  the     different photometric bands, respectively black for
$B$,   blu   for  $V$, red for $R$, and yellow for $I$. Also in this case
we    observed a linear     increase  in   size with   a  mean velocity of
$(14.6\,\pm\,0.3)$ arcsec/day ($0.200\,\pm\,0.004\,$ km/sec).   Repeating
the            calculation           separately       for  the  different
bands         gave         the        result in Tab.~\ref{Table:velcloud}.
These          mean           values            and     their      errors
were obtained      using       a      weighted    least square algorithm.
The different bands allow to explore different layers of  the   expanding
cloud, more  external  in   the  $B$,  and     deeper   in   the   redder
bands. Thus, this result can be  interpreted  as   the   evidence   of  a
radial velocity gradient in the expanding cloud.      For each couple  of
adiacent            bands       we       considered the ratio between the
difference of the  expansion velocities reported in Tab.~\ref{Table:velcloud},
and the difference in the correspondent $\sigma$ values given   by    the
least                     square                                  fitting
models for the night 2/11/2007,  for which we obtained the largest separation among
the expanding shells. Taking the mean of these values we obtained an estimate
of        the      radial     velocity       gradient during that night equal to
$\rm (0.3\,\pm\,0.2)\,10^{-5}\,sec^{-1}$    which     means an increase of
around $0.3\,$cm/sec every km going from the center to the surface of
the coma.

All   these     kinematic    observations  could    be    explained in the
context of an explosive scenario. While, as a consequence of the explosion,
a part of the  cometary nucleus was disintegrated   and  the material was
outflowed in all directions resulting in the sperically symmetric expanding
coma, the survived nucleus received a kick  separating from the coma's
center with the observed projected velocity.      The  expansion velocity
of a particle of  radius    $a$ during an explosive event is proportional
to $a^{-1}$   (see    e.g.  the     discussion    in    Tozzi et al. 2007)
implying   a        higher  expansion velocity for    the        smaller,
less                 massive             particles                   with
respect            to          the          larger and more massive ones,
which               resulted             in        the observed  velocity
gradient              in             the            expanding      cloud.

The formation  of   spherical   expanding envelopes around the   cometary
nucleus of \object{17P/Holmes} was observed for both the events  occurred
in  November 1892  and  in              January                     1893.
Bobrovnikoff (1943)           compared           the         observations
obtained     by     different     authors   between   16-22 January 1893,
deriving   a   uniform expansion velocity of $(0.54\,\pm\,0.03)\,$km/sec,
thus           similar             to       what       found        here.
The     sudden   brightness      increases,           the   formation of
spherical  dust  envelopes   with  similar  kinematic properties,     the
correspondence in the orbital phase at the instant of the major outbursts
noted in Sect.~\ref{sect:introduction}, point towards a common  mechanism
at the base of the observed phenomena. Moreover, the distance    from the
ecliptic              plane           was in both cases around $0.8\,$AU.
Given                that,  although it    cannot be completely excluded,
it                   seems               unlikely                    that
the above mention explosive mechanism was prompt by an asteroidal impact,
beeing thus more probable an internal instability process .

\begin{table}
\caption{The derived expansion velocities  in  arcsec/day  along  with  their
uncertainties    for   the   expanding   dust  cloud in the different bands.
In parenthesis we report the correspondent values in km/sec.}
\label{Table:velcloud}
\centering
\begin{tabular}{c c c c}
\hline\hline
Filter & Expansion vel. & $\pm$ & err \\
\hline
 $B$ & 16.2(0.222) & & 0.5(0.007) \\
 $V$ & 14.0(0.193) & & 0.6(0.009) \\
 $R$ & 13.5(0.185) & & 0.2(0.003) \\
 $I$ & 11.9(0.163) & & 0.4(0.005) \\
\hline
\end{tabular}
\end{table}

%


\section{The ejected mass estimate}
\label{sect:mass}

\begin{figure}
   \centering
   \includegraphics[width=0.49\columnwidth]{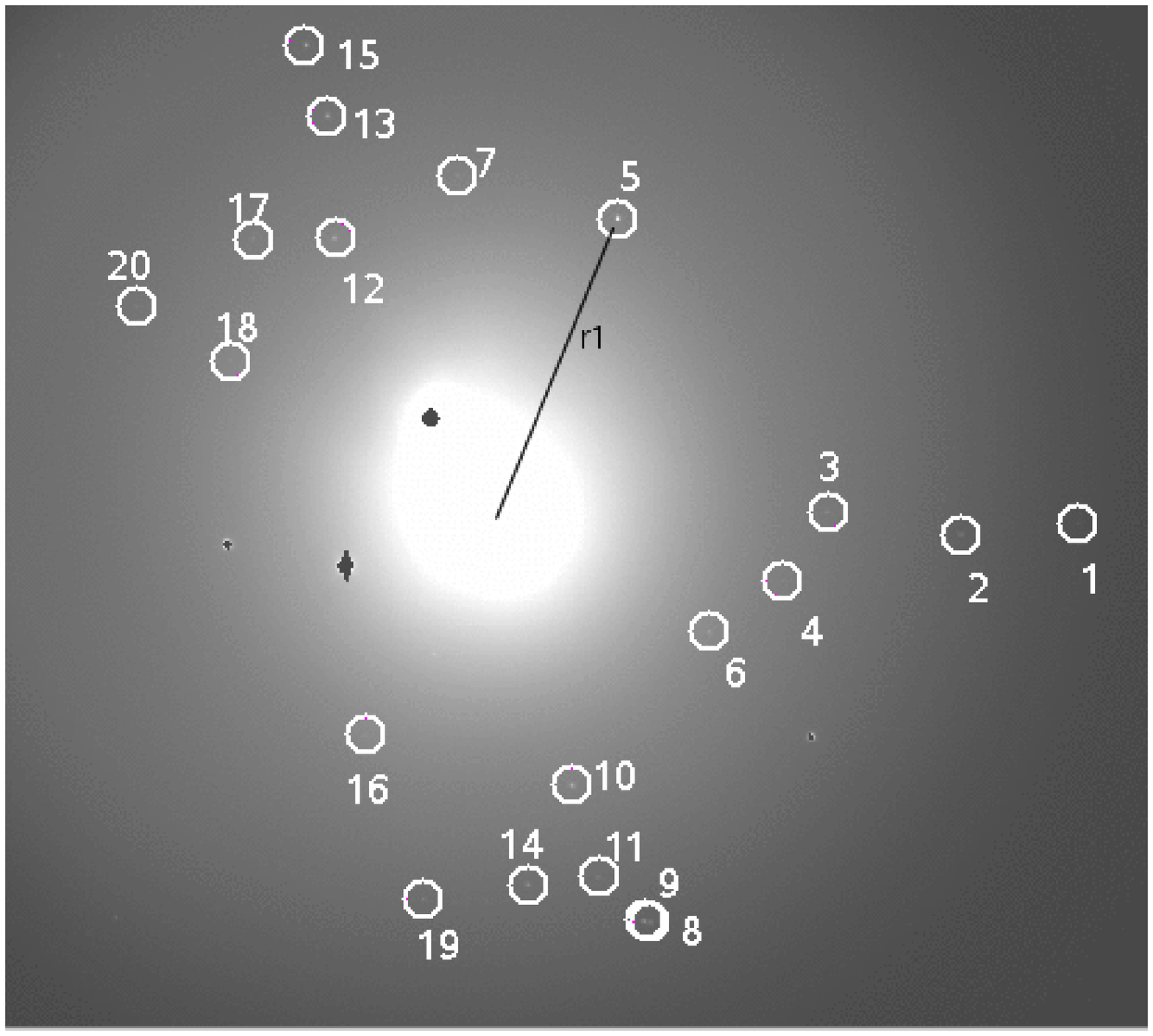}
   \includegraphics[width=0.497\columnwidth]{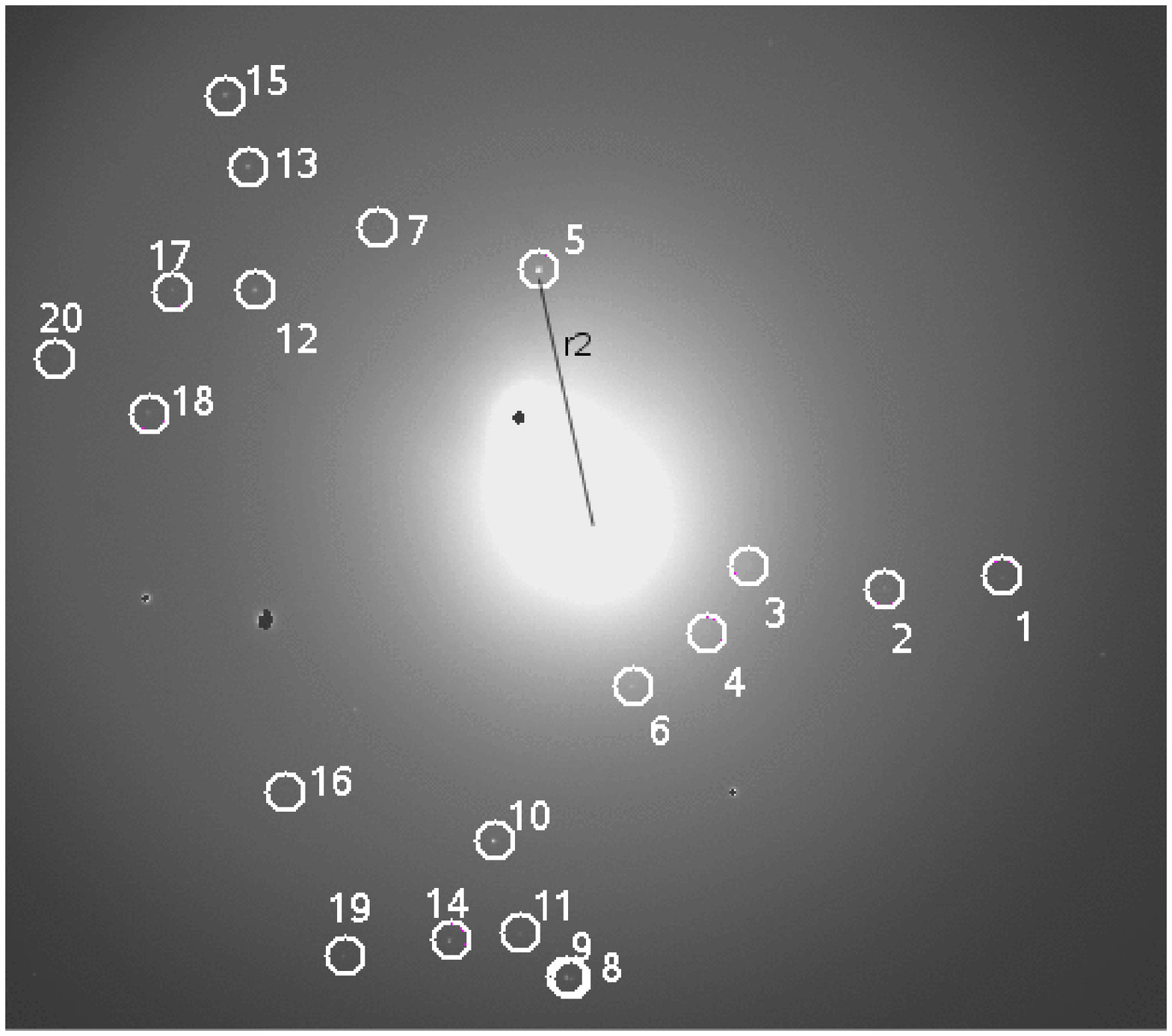}
   \includegraphics[width=0.8\columnwidth]{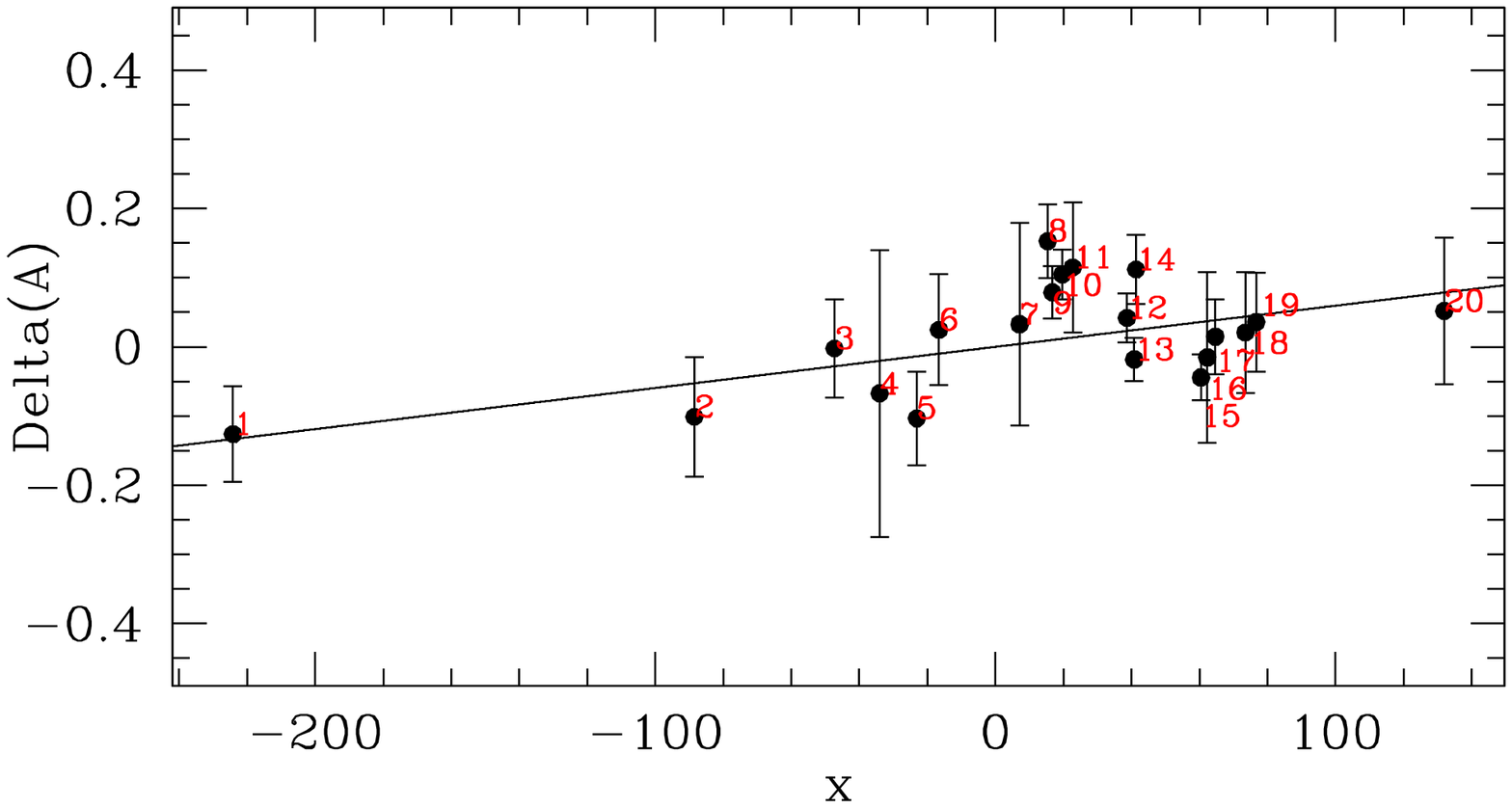}
   \caption{Top: spatial    distribution   of   the   $20$ background stars
            used    in      Sect.~\ref{sect:mass} to probe the differential
            extinction produced by the coma, for the two images acquired on
            October 28, 2007. The image on the left was taken $2$ hours before
            the                    right                               one.
	    The                   selected                stars         are
            indicated      by    white circles. The  associated  numeration
            corresponds to that one in     the       lower           panel,
            allowing    to     relate   each star   to   the  correspondent
            measured              differential                   extinction.
            Bright           sources               were           saturated
            (black regions), as well as the center of the cometary nucleus.
            The       black      lines indicate the distances, referred  in
            Sect.~\ref{sect:mass}   as   $\rm r_1$      and $\rm r_2$,   of
            one       star       with     respect             to       the
            estimated     coma's                                   centers.
            Bottom: the correlation between the differential extinctions in
            the $R$ band   ($\rm \Delta\,A$), and the $\rm x$     parameter
            (see text for details).
            }
              \label{Fig:massimages}%
    \end{figure}

In the following we provide an estimate of the ejected  dust mass during the
outburst,    through    the   extinction   produced by the dust cloud on the
surrounding      background       stars.      We  selected two well exposed,
good     seeing    images    taken   on  October 28, 2007 in the $R$ filter.
These        images           had        the           largest          time
separation   ($2$ hours)   in  the whole dataset, among equal filter  images
acquired     during       the       same         observing            night.
Thus,        they           provided         the          largest    apparent
motion    of    the    cloud   on the sky plane. At the same time,      this
temporal       difference   is   small enough to avoid the expansion of  the
cloud to change                            significantly    the   extinction
map,             allowing           an                homogeneous comparison
of     the    two       images.        Finally,       day     28   was as close
to     the  outburst as to allow the dust cloud to fit well inside our field of
view.
We          performed         PSF          fitting        photometry    with
DAOPHOT/ALLSTAR              (Stetson 1987)         and             rejected
all           the             stars              with         $\chi>1.5$ and
absolute         $sharp$             values          bigger   than      $1$.
These parameters  and selection criteria     allowed  to exclude those objects
for                                                                    which
a   reliable fit of the stellar model couldn't be performed,   because close
to saturation,    bad  pixels,       and to reject  non-stellar      sources
like           cosmic         rays          or      background     galaxies,
spikes                    of            saturated          stars        etc.
The sky background in these images was estimated locally around each star,
in  an  annular region comprised between $1.5$ and $3.5$ arcsec from the stars'
centroids.  This region was selected after performing different tests looking
at the best subtraction of the analized stars. The modal value of the   pixel
counts inside the selected  annular region  was considered as the estimate of
the      sky       background.   The photometry was extracted in a circular
region centered on the stars' centroids with radius equal to $1.3\,$arcsec, which
corresponded  to the seeing value in both the images. The derived magnitudes
were  then   reported to $1$ airmass and $1\,$ sec of exposure time for both the
images.
  We then consider all the common
objects       with     radial     distance  from  the  center  of  the comet
$>25$ arcsec,      and     with       radial   distance     from  the center
of       the          cloud        $<3$       arcmin     in both the images.
The      inner        limit          was        necessary    to    avoid the
photometry of the stars to be biased by the luminous  core of the comet. The
outer     limit       was         chosen  to avoid detector border  effects,
accounting   for the position of the center of the cloud in both the images.
At     the      end,       we        obtained       a       catalog       of
$20$       stars      spread      inside   the analized region of the cloud,
see Fig.~\ref{Fig:massimages}.

In    order    to     demonstrate    the    radial    dependence   of   the
extinction                    from               the center  of  the  cloud,
we                 assumed             a            uniform and homogeneous
spherically                   symmetric               mass    distribution.
Thus,          being            $\rm r$            the            position
of     one      star    with      respect     to      the cloud's    center,
we              can                    express the optical        depth  of
the cloud at  the  distance $\rm r$ as:

\begin{equation}
\rm \tau(r)=\sigma_d\,n\,2\sqrt{\rm R^2-r^2}
\end{equation}

\noindent
where      $\rm 2\sqrt{R^2-r^2}$           is        the              lenght
of    the      segment        of            sphere       with      projected
distance $\rm r$,       with        respect      to    the cloud's center, $\rm R$
is   the    radius   of   the  cloud, $\rm \sigma_d$ is the cross section of
the dust grains in a given observing band, and $\rm n$ is the number density
of the dust particles. Considering two different positions    $\rm r_1$  and
$\rm r_2$   of    a   generic    star   in the first and in the second image
with                     respect             the           cloud's      center
we can express the differential extinction $\rm \Delta\,A$,  produced by the
cloud in the two positions as:

\begin{equation}
\rm \Delta\,A\,\,\,=\,\,\,m_1\,-\,m_2\,=\,-2.5\,\log\left(\frac{I_1}{I_2}\right)=\,-2.5\,log\left[\frac{e^{-\sigma_d\,n\,2\,\sqrt{\rm R^2-r_1^2}}}{e^{-\sigma_d\,n\,2\,\sqrt{\rm R^2-r_2^2}}}\right]
\end{equation}

which implies:

\begin{displaymath}
\rm \Delta\,A\,\,\,\,=\,\,\,(2.5\,\sigma_d\,n\,2\,\log\,e)\,\Big(\sqrt{\rm R^2-r_1^2}-\sqrt{\rm R^2-r_2^2}\Big)\,\,\,\,=
\end{displaymath}

\begin{equation}
\rm\,\,\,\,\,\,\,\,\,\,\,\,\,\,\,=\,\,\,\,\alpha\,\,\,\,\Big(\sqrt{\rm R^2-r_1^2}-\sqrt{\rm R^2-r_2^2}\Big)
\end{equation}

\noindent
where    $\rm m_1$,$\rm I_1$     and     $\rm m_2$,$\rm I_2$ are the $observed$
magnitudes (intensities)   of  the star at the position $\rm r_1$ and $\rm r_2$
respectively.    As    easily     recognized,      this    model     satisfies
the     required          simmetry         conditions      with        respect
to        $\rm r_1$         and              $\rm r_2$.     Actually,       if
$\rm r_1$   is  equal to $\rm r_2$, the differential extinction should be zero.
Morever, if  we  consider  two stars with initial $\rm r_1$, $\rm r^{'}_1$, and
and   final      $\rm r_2$,       $\rm r^{'}_2$      positions    respectively,
that                   satisfy                  the                  condition
$\rm r_1=r^{'}_2$        and       $\rm r_2=r^{'}_1$,           the   resulting
differential            extinctions          for                           the
two         stars          should         be        equal and opposite in sign.
In         particular               $\rm \Delta\,A$                     should
be positive when $\rm r_1<r_2$, (the star is moving away from the cloud center,
and           thus            is           less       extincted  in  $\rm r_2$
with       respect         to          $\rm r_1$),         and        negative
when       $\rm r_1>r_2$      (the        star       is   going    towards the
cloud              center).           This            implies             that
the      angular      coefficient  $\rm \alpha$, in Eq.~3, should be positive.
The    value        of       $\rm \alpha$         gives   the       extinction
for         unitary        length        in       the    given observing band.
Therefore,      we    fit  to the observed differential extinctions the model:

\begin{equation}
\rm \Delta\,A\,=\,\alpha\,x\,+\beta
\end{equation}

\noindent
where                       $\rm x=\sqrt{\rm R^2-r_1^2}-\sqrt{\rm R^2-r_2^2}$,
and      the          $\rm \beta$   coefficient was included to   account for
residual constant zero points between  the  two images.
The result, after subtracting  the
constant   zero   point, is shown in the bottom        panel               of
Fig.~\ref{Fig:massimages}.                  We         found     a   positive
correlation   between      the    differential   extinction  $\rm \Delta\,A$,
and     the      $\rm x$      parameter,       as     expected.   The derived
value        of        the        $\rm \alpha$      coefficient           was
$(0.0006\,\pm\,0.0002)\,$mag/pix,          where          a             pixel
corresponds               to                   around               $592\,$km
at           the          distance        of     the    comet. This value has
been     obtained    assuming      a  radius $\rm R$ for the cloud of $3\,$arcmin
($\rm \sim\,3.5\,\rm \sigma$    for  that night for the luminous distribution
analized in Sect.~\ref{sect:cinematic}). Increasing the radius to $4\,$arcmin
 ($\rm \sim\,5\,\rm \sigma$)           implies          a         value    of
$\rm \alpha\,=\,(0.0012\,\pm\,0.0004)\,$mag/pix.

The    most        important    assumptions        of           the      model
are the dust cloud's spherical geometry    and   the homogeneous   and uniform
mass        distribution.                   The     geometry     of        the
cloud         was         well       constrained       by     the observations
and     the      analysis       presented       in      the           previous
Sections. As for the     second                    hypothesis,              it
allowed      to      express         the optical depth in a straightforward and
convenient     way,      as        shown       in            Eq.~1,        and
ultimately      implies      the        linear                      dependence
of     the         observed       differential  extinctions on the geometrical
factor $x$. A           strong           deviation      from   that assumption
would   imply    a     strong      deviation from the linear model prediction,
which        is         not         observed      (Fig.~\ref{Fig:massimages}).
The                        scatter           in            that       relation
($\sim0.06\,$ mag)            is                 larger                   than
the  scatter       derived        from       the    calibration             of
the Landolt standard field discussed in Sect.~\ref{sect:observations}     ($\sim0.02\,$ mag).
Otherwise,             in             the                           science
images the surface brightness of the coma  determined a background around $10$
times larger than in the Landolt images,   which explains    the factor of $3$
increase in the scatter.
The   radial       velocity gradient of the expanding material
discussed in Sect.~\ref{sect:cinematic}        points against the uniform mass
distribution             hypothesis.                Otherwise,              as
shown           in   the right panel of Fig.~\ref{Fig:velocities},     during
night $28$ the  expanding shells were certainly more concentrated     together
than                  in         the         later       evolved        phases
beeing         that           night          closer      to      the outburst.
In conclusion,    it is reasonable to believe that in the chosen night the
material was not far from beeing uniformly distributed and homogeneously mixed
and           we             considered     this        hypothesis   as a good
approximation    of     the      structural properties of the observed   coma.
Therefore, the dominant factor that explains the observations is related to
the cloud's geometrical structure, in the sense that the observed differential
extinctions were determined by the variable quantity of mass over the cloud's
different line of sights probed by the background stars (directly implied by its
spherical structure). This is overall demonstrated by the observed linear dependence
of the differential extinction from the geometrical factor x, and by the
positive value of the $\alpha$ coefficient, also predicted by the model.

To        derive        the      coma's        mass        we       considered
spherical dust grains with a mean density   $\rm \rho_d$  ($\rm =2.5\,g/cm^3$)
and            a         'typical'             dimension             $\rm r_d$
(and              thus              mass    $\rm m_d=4/3\,\pi\,r_d^3\,\rho_d$).
Thanks          to      the definition  of the $\rm \alpha$ parameter in Eq.3,
the mass of the cloud $\rm M$     can     be expressed  through the following
formula:

\begin{displaymath}
\rm M=\frac{4}{3}\,\pi\,R^3\,m_d\,n=\frac{4}{3}\,\pi\,R^3\,\left(\frac{4}{3}\,\pi\,r_{d}^3\,\rho_d\right)\left(\frac{\alpha}{5\,\sigma_d\,\log\,e}\right)=
\end{displaymath}

\begin{equation}
\rm \,\,\,\,\,\,\,\,=\frac{16\,\pi}{45\,\log\,e}\,R^3\,r_d\,\rho_d\,\alpha,\simeq\,R^3\,r_d\,\rho_d\,\alpha
\end{equation}

\noindent
where    the                 grain's                             cross-section
was taken equal to the geometrical   cross-section $\rm \sigma_d\,=\,\pi\,r_d^{2}$
and            the         numerical      factor         is       $\sim\,2.6$.
We                                                              varied     the
value of $\rm r_d$    inside    a range of  characteristic   grain dimensions,
$\rm 0.005-1\,\mu$m         (Mathis et al. 1977),   and   the dimension of the
cloud                $\rm R$        between                     $3-4\,$arcmin.
The          derived      estimate  for the coma's  mass was comprised between
$10^{12}-10^{14}$kg. Snodgrass et al.~(2006) provided the most recent    values
for  this comet's              dimension        and               density.  In
particular,   from  their time-series photometry they obtained a value for the
effective radius of  the nucleus (Russel 1916) of $\rm (1.62\,\pm\,0.01)\,km$.
Even if accurate, this    estimate  is    based  on the   assumption that the
geometrical            albedo          was            equal to $\rm A_R=0.04$.
Furthermore,     they       derived        a comet's minimum  density equal to
$\rm \rho\,=\,0.09\,g/cm^3$. Assuming          thus         a       range of
possible                    densities           ($\rm \rho\,=\,0.1-1\,g/cm^3$)
and                                                                    albedos
($\rm A_R\,=\,0.01-0.1$) and         using               Eq.~$\rm (1)$      in
Snodgrass et al.~(2006)                  to                   derive       the
correspondent effective radii, we obtained a total nuclear mass      comprised
between $10^{12}-10^{14}$kg. Given the uncertainty range in our mass estimate we
concluded that the probable value of the expanding coma's       mass resulting
from the outburst     event was comprised between $10^{-2}-1$ comet's    mass.

There    are      different       factors       that  could affect our result.
Dust         grains       in      cometary    ejecta typically span a range of
different dimensions and optical properties (see e.g. Lisse et al 2007, Tozzi et
al. 2007).   Anyway, both the assumption on the uniform and homogeneous   mass
distribution, and the lack of specific observations able to constrain the dust
population characteristics for this  particular object, justify our simplified
approach to  summarize the cloud's grain content    assuming a 'typical' grain
dimension and the correspondent geometrical cross-section.
Another        possible      drawback    in   our coma's    mass      estimate
regards       the     pre-outburst     activity     of        the       comet.
During              the      most           recent                observations
of the  comet obtained before the outburst (Snodgrass et al.~2006)  the object
appeared to be inactive,     although  the  heliocentric distance at that time
was $\sim4.66\,$AU,    whereas    at     the  outburst         $\sim2.44\,$AU.
Moreover,   it seems   probable    that the explosive event described in this
work largely overcame the common activity of the comet. Finally, the stars used
to                  probe                   the                   differential
extinction produced by the coma (Fig.~\ref{Fig:massimages}),   were
well     distributed      inside   the     coma,     avoiding the region close
to the comet's nucleus, and thus reflecting more closely  the contribution to the
differential    extinction of the material coming directly from the explosion.

Despite                the              above        mentioned  approximations,
the              result             presented         in                 Eq.~5,
has           the            advantage         to         be            simple,
to enlight                         the                              dependence
among     the  total mass of the cloud,          the  cloud's     geometrical,
composition                                                         properties
and the observed differential extinctions, and to provide      an     estimate
for the expanding coma's  mass  which  points   towards    a        consistent
disintegration      phenomenon,              as            suggested        by
the                        large                       outburst         event.


\section{Conclusions}
\label{sect:conclusions}

In    this  Letter we analized the early phases of the outburst that 
comet  \object{17P/Holmes} underwent   on    October 24, 2007.    We acquired    $BVRI$
photometric      images      at       the      Wendelstein $0.8\,$m telescope,
between 26/10/2007 and 20/11/2007.  We   observed         a        spherically
simmetric              dust               cloud                         moving
away    from      the comet nucleus with a mean projected constant velocity of
$(9.87\,\pm\,0.07)$    arcsec/day   ($0.135\,\pm\,0.001$ Km/sec),  while   the
dust       cloud      was     expanding     with    a   mean constant velocity
of $(14.6\,\pm\,0.3)$       arcsec/day         ($0.200\,\pm\,0.004\,$ Km/sec).
These         results        are        in       agreement     with the ones
obtained       during        past        outbursts      of      this    comet.
Evidence   of  a gradient in the expansion velocity of the dust cloud was also
found   with the velocity increasing towards the external regions.    Finally,
performing     differential     photometry   on      background stars occulted
by      the   moving  cloud,  and assuming a uniform and homogeneous spherical
mass distribution,     we      derived      for     the coma's mass a value of
$10^{12}-10^{14}$kg,          around                   $10^{-2}-1$ comet's mass.
We    interpreted  our observations in the context of an explosive event,
probably caused by some internal instability processes,  rather than  by    an
asteroidal         body's             impact.
Various               mechanisms   have     been    proposed    in  the   past
to explain    comets'    splitting,    involving    tidal,  thermal
and    rotational  forces (Sekanina 1997).   These        processes      could
have  played an important    role      for the         event        discussed here,
although  some  specific  characteristics allow to consider it as more peculiar.
For example, generally  the separation velocities of the  splitting  components
are of the order of a few $\rm m/s$, while   in this case we found a projected
relative velocity around $2$ orders of magnitude  larger. The outburst  itself
represents the largest apparent brightness increase ever observed for a  comet.
In    conclusion,    we     underline  the importance to consider
other            observational  results   in   order to accurately characterize
this             event               and        to     provide   a more insight
view on         its           enigmatic and still not well  understood  nature.

{\bf
\begin{acknowledgements}
We warmly thanks the anonymous Referee for the helpful comments and
suggestions allowing a significant improvment of the Letter.
\end{acknowledgements}
}

\end{document}